\documentclass[aps,prd,twocolumn,showpacs,amsmath,amssymb]{revtex4-1}
\usepackage{amsmath} \usepackage{graphicx} \usepackage{subfigure}
\usepackage{epstopdf} \usepackage{color} \usepackage{multirow}
\usepackage{setspace} \usepackage{overpic} \usepackage{amssymb}
\usepackage[bookmarksnumbered, pdfstartview=FitH,colorlinks,urlcolor=blue, citecolor=blue,linkcolor=blue] {hyperref}
\usepackage{lineno}
\usepackage{bm}
\usepackage{rotating}
\usepackage{xcolor}
\usepackage{makecell}
\usepackage{mathtext}
\usepackage{mathrsfs}
\usepackage{overpic}
\usepackage[T1]{fontenc}
\usepackage{lmodern}
\usepackage[utf8]{inputenc}
\usepackage{float}
\usepackage{cuted}

\newcommand{\PreserveBackslash}[1]{\let\temp=\\#1\let\\=\temp}
\newcolumntype{C}[1]{>{\PreserveBackslash\centering}p{#1}}
\newcolumntype{R}[1]{>{\PreserveBackslash\raggedleft}p{#1}}
\newcolumntype{L}[1]{>{\PreserveBackslash\raggedright}p{#1}}

\let\oldequation\equation
\let\oldendequation\endequation
\renewenvironment{equation}{\linenomathNonumbers\oldequation}{\oldendequation\endlinenomath}

\begin{document}

\graphicspath{{figure/}}
\DeclareGraphicsExtensions{.eps,.png,.ps}

\title{\boldmath Improved search for $\psi(3770) \to \gamma \eta_{c}(1S, 2S)$ radiative transitions}
\author{
  \begin{small}
    \begin{center}
M.~Ablikim$^{1}$\BESIIIorcid{0000-0002-3935-619X},
M.~N.~Achasov$^{4,c}$\BESIIIorcid{0000-0002-9400-8622},
P.~Adlarson$^{85}$\BESIIIorcid{0000-0001-6280-3851},
X.~C.~Ai$^{91}$\BESIIIorcid{0000-0003-3856-2415},
C.~S.~Akondi$^{32A,32B}$\BESIIIorcid{0000-0001-6303-5217},
R.~Aliberti$^{40}$\BESIIIorcid{0000-0003-3500-4012},
A.~Amoroso$^{84A,84C}$\BESIIIorcid{0000-0002-3095-8610},
Q.~An$^{80,67,\dagger}$,
M.~S.~Anderson$^{40}$\BESIIIorcid{0009-0008-1550-2632},
Y.~Bai$^{65}$\BESIIIorcid{0000-0001-6593-5665}
% ...
\\
\vspace{0.2cm}
(BESIII Collaboration)\\
\vspace{0.2cm} {\it
$^{1}$ Institute of High Energy Physics, Beijing 100049, People's Republic of China\\
% ...
$^{4}$ Budker Institute of Nuclear Physics SB RAS (BINP), Novosibirsk 630090, Russia\\
% ...
$^{32}$ INFN Sezione di Ferrara, (A)INFN Sezione di Ferrara, I-44122, Ferrara, Italy; (B)University of Ferrara, I-44122, Ferrara, Italy\\
% ...
$^{40}$ Johannes Gutenberg University of Mainz, Johann-Joachim-Becher-Weg 45, D-55099 Mainz, Germany\\
% ...
$^{65}$ Southeast University, Nanjing 211100, People's Republic of China\\
% ...
$^{67}$ State Key Laboratory of Particle Detection and Electronics, Beijing 100049, Hefei 230026, People's Republic of China\\
% ...
$^{80}$ University of Science and Technology of China, Hefei 230026, People's Republic of China\\
% ...
$^{84}$ University of Turin and INFN, (A)University of Turin, I-10125, Turin, Italy; (B)University of Eastern Piedmont, I-15121, Alessandria, Italy; (C)INFN, I-10125, Turin, Italy\\
$^{85}$ Uppsala University, Box 516, SE-75120 Uppsala, Sweden\\
% ...
$^{91}$ Zhengzhou University, Zhengzhou 450001, People's Republic of China\\
\vspace{0.2cm}
$^{\dagger}$ Deceased\\
% ...
$^{c}$ Also at the Novosibirsk State University, Novosibirsk, 630090, Russia\\
% ...
      }\end{center}
    \vspace{0.4cm}
\end{small}
}
\affiliation{}


\begin{abstract}
Based on an integrated luminosity of $20.3~\mathrm{fb}^{-1}$ of $e^{+}e^{-}$ annihilation data collected at a center-of-mass energy of $3.773~\rm{GeV}$ with the BESIII detector operating at the BEPCII collider, an improved search 
for the radiative transitions $\psi(3770) \to \gamma \eta_{c}(1S, 2S)$ is performed using
the hadronic decays $\eta_{c}(1S, 2S) \to K^{0}_{S} K^{\pm} \pi^{\mp}$.
No significant signal is observed. 
The corresponding 90$\%$ confidence level upper limits on the product branching fractions are set to be $5.0 \times 10^{-6}$ for the $\eta_{c}(1S)$ transition and $3.7 \times 10^{-6}$ for the $\eta_{c}(2S)$ transition. The 90$\%$ confidence level upper limits on the partial decay widths are also reported to be $\Gamma(\psi(3770) \to \gamma \eta_{c}(1S)) < 5.5$ keV and $\Gamma(\psi(3770) \to \gamma \eta_{c}(2S)) < 29.4~\rm{keV}$. With about seven times larger integrated luminosity than used previously, these results lower the upper limits by approximately a factor of three and two for the $\eta_{c}(1S)$ and $\eta_{c}(2S)$ transitions, respectively.
\end{abstract}


\newcommand{\BESIIIorcid}[1]{\href{https://orcid.org/#1}{\hspace*{0.1em}\raisebox{-0.45ex}{\includegraphics[width=1em]{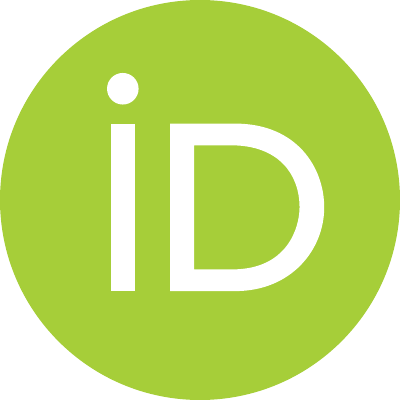}}}} 

\maketitle
\section{Introduction}
The $\psi(3770)$ resonance is the lightest $J^{PC} = 1^{--}~c\bar{c}$ state lying above the open charm threshold and is generally considered to be the $1^{3}D_{1}$ state of charmonium with a small $2^{3}S_{1}$ component~\cite{Ref_admixture}. A comprehensive understanding of the properties of the $\psi(3770)$ meson remains a long-standing challenge. Early expectations were that the $\psi(3770)$ would decay almost entirely to $D\bar{D}$ final states~\cite{Ref_obse1,Ref_obse2}. However, the BES Collaboration reported a large inclusive non-$D\bar{D}$ branching fraction of $(14.7 \pm 3.2)\%$ under the hypothesis that a single $\psi(3770)$ resonance exists in the center-of-mass energy ($\sqrt{s}$) range between 3.70 and 3.87 GeV~\cite{Ref_method1,Ref_method2,Ref_method3,Ref_method4}. In contrast, the CLEO Collaboration reported a non-$D\bar{D}$ branching fraction of $(-3.3 \pm 1.4 ^{+6.6}_{-4.8})\%$ after accounting for the interference between resonant $\psi(3770)$ production and non-resonant $e^{+}e^{-}$ annihilation~\cite{Ref_cleobf}. These discrepancies highlight the importance of using exclusive decay measurements to resolve this puzzle. According to the Particle Data Group (PDG)~\cite{Ref_pdg}, nine exclusive non-$D\bar{D}$ decay modes of the $\psi(3770)$ meson have been observed, including $J/\psi\pi^{+}\pi^{-}$~\cite{Ref_CLEO:2005zky}, $J/\psi\pi^{0}\pi^{0}$~\cite{Ref_CLEO:2005zky}, $J/\psi\eta$~\cite{Ref_BESIII:2022yoo}, $\phi\eta$~\cite{Ref_CLEO:2005zrs}, $K^{0}_{S}K^{0}_{L}$~\cite{Ref_BESIII:2023zsk}, $\mathit{\Xi}^{-}\bar{\mathit{\Xi}}^{+}$~\cite{Ref_BESIII:2023rse}, $e^{+}e^{-}$~\cite{Ref_BES:2006piq}, and $\gamma\chi_{c0, c1}$~\cite{Ref_BESIII:2015cby}.  The sum of the observed non-$D\bar{D}$ exclusive components is only about 1.3$\%$. Among the observed decays, the radiative transitions $\psi(3770) \to \gamma\chi_{c0,c1}$ have been found to have non-negligible branching fractions.  This motivates the further exploration of other radiative transitions, such as $\psi(3770) \to \gamma\eta_c(1S,2S)$, which can be used to test non-perturbative mechanisms~\cite{Ref_IML1,Ref_IML2,Ref_IML3}.

If the $\psi(3770)$ meson were a pure $1^{3}D_{1}$ state, the magnetic dipole ($M1$) transitions $\psi(3770) \to \gamma \eta_{c}(1S, 2S)$ would be forbidden. However, the radiative photon has a finite energy, which allows contributions from higher multipoles beyond the leading order~\cite{Ref_iml}. Intermediate meson loops (IML) play an important role in exclusive transitions, especially when the initial state mass is close to the open channel threshold~\cite{Ref_IML1,Ref_IML2,Ref_IML3}. By considering contributions from the IML mechanism, the partial decay widths are calculated to be $\Gamma(\psi(3770) \to \gamma \eta_{c}(1S)) = (17.14^{+22.9}_{-12.03})$ keV and $\Gamma(\psi(3770) \to \gamma \eta_{c}(2S)) = (1.82^{+1.95}_{-1.19}) $ keV~\cite{Ref_iml}, with corresponding branching fractions $\mathcal{B}(\psi(3770) \to \gamma \eta_{c}(1S)) = (6.3^{+8.4}_{-4.4}) \times 10^{-4}$ and $\mathcal{B}(\psi(3770) \to \gamma \eta_{c}(2S)) = (6.7^{+7.2}_{-4.4}) \times 10^{-5}$, calculated using $\Gamma_{\psi(3770)} = 27.2 \pm 1.0$ MeV~\cite{Ref_pdg}. These values are of the same order as Lattice Quantum Chromodynamics (LQCD) predictions~\cite{Ref_lqcd}, suggesting that interference between the $M1$ transition amplitude in the pure $c\bar{c}$ quark model and the contributions from IML involving open-charm channels may play an important role in $\psi(3770)$ radiative decays~\cite{Ref_iml}.

In addition, calculations of charmonium transition rates generally apply small relativistic corrections and have been treated using non-relativistic models. However, Ref.~\cite{Ref_psi2} indicates that relativistic corrections, including the Electric Quadrupole ($E2$) and Magnetic Octupole ($M3$), can be comparable to the non-relativistic leading-order ($M1$) contribution. Ref.~\cite{Ref_LRC} employs the relativistic Bethe-Salpeter (BS) equation method to calculate the electromagnetic radiative decays of heavy quarkonium, taking the $M1$ transition as the leading-order contribution and focusing on the relativistic corrections from higher-order multipole transitions. The large relativistic corrections lead to significantly smaller predictions: $\Gamma(\psi(3770) \to \gamma \eta_{c}(1S)) < 1.54$ keV and $\Gamma(\psi(3770) \to \gamma \eta_{c}(2S)) < 0.0233$ keV. Further experimental measurements of the $\psi(3770) \to \gamma \eta_{c}(1S, 2S)$ transition rates will be helpful for testing theoretical predictions and constraining IML contributions.

The BESIII Collaboration previously searched for the radiative transitions $\psi(3770) \to \gamma \eta_{c}(1S, 2S)$ using a data sample with an integrated luminosity of $2.92~\mathrm{fb}^{-1}$ taken at the $\psi(3770)$ resonance~\cite{Ref_3fb}. Since no significant signals were observed, 90$\%$ confidence level (C.L.) upper limits on $\Gamma(\psi(3770) \to \gamma \eta_{c}(1S, 2S))$ were set to be 18.6 keV and 54.8 keV, respectively. Using a significantly larger $\psi(3770)$ resonance data sample provides an opportunity to improve these measurements and verify the different theoretical predictions.

In this article, improved measurements of the $\psi(3770) \to \gamma \eta_{c}(1S, 2S)$ transition rates are presented using a $\psi(3770)$ resonance data sample with an integrated luminosity of $20.3~\mathrm{fb}^{-1}$. The $\eta_{c}(1S, 2S)$ are reconstructed via $\eta_{c}(1S, 2S) \to K^{0}_{S} K^{\pm} \pi^{\mp}$, which benefits from a large branching fraction, high reconstruction efficiency, and a low combinatorial background level.  Other prominent decay channels of the $\eta_{c}(1S, 2S)$ either have much lower reconstruction efficiencies or suffer from  high background levels. Using this larger data sample, we significantly improve the 90$\%$ C.L. upper limits on $\Gamma(\psi(3770) \to \gamma \eta_{c}(1S, 2S))$, and subsequently compare our results with theoretical predictions.

\section{BESIII DETECTOR AND MONTE CARLO SIMULATION}
\label{sec:BES}

The BESIII detector is a magnetic spectrometer~\cite{besiii,besiii2} operating at the BEPCII collider~\cite{bepcii}. The BESIII detector records symmetric $e^{+}e^{-}$ collisions in the $\sqrt{s}$ range from 1.84 to 4.95 GeV, with a peak luminosity of $1.1 \times 10^{33}~\rm{cm}^{-2}\rm{s}^{-1}$ achieved at $\sqrt{s} = 3.773$ GeV. BESIII has collected large data samples in this energy region~\cite{data1,data2}. The cylindrical core of the BESIII detector covers $93\%$ of the $4\pi$ solid angle and consists of a helium-based multilayer drift chamber (MDC), a plastic scintillator time-of-flight system (TOF), and a CsI (Tl) electromagnetic calorimeter (EMC), which are all enclosed in a superconducting solenoidal magnet, providing a 1.0~T magnetic field. The solenoid is supported by an octagonal flux-return yoke, with resistive plate chamber muon identifier modules interleaved with steel. The charged-particle momentum resolution at $1~{\rm GeV}/c$ is $0.5\%$, and the $\textrm{d}E/\textrm{d}x$ resolution is $6\%$ for the electrons from Bhabha scattering. The EMC measures photon energies with a resolution of $2.5\%$ ($5\%$) at $1$~GeV in the barrel (end cap) region. The time resolution of the TOF barrel section is 68~ps, while that of the end cap section is 110~ps. The end cap TOF system was upgraded in 2015 with multi-gap resistive plate chamber technology, providing a time resolution of 60~ps~\cite{etof1, etof2, etof3}, which benefits about $86\%$ of the data used in this analysis.

Simulated Monte Carlo (MC) samples are produced with the {\sc geant4}-based~\cite{G4} software, which models the experimental conditions, including the electron-positron collision, the decays of the particles, and the response of the detector. The simulation models the beam energy spread and initial state radiation (ISR) in the $e^{+}e^{-}$ annihilations with the generator {\sc kkmc}~\cite{kkmc}. To investigate possible background contamination, inclusive MC samples for different processes are generated, which include $e^{+}e^{-} \to \psi(3770) \to D^{0}\bar{D}^{0}$, $e^{+}e^{-} \to \psi(3770) \to D^{+}D^{-}$, $e^{+}e^{-} \to \psi(3770) \to$ non-$D\bar{D}$, $e^{+}e^{-} \to q\bar{q}~(q = u, d, s)$, $e^{+}e^{-} \to \gamma_{\mathrm{ISR}}\psi(3686)$, and $e^{+}e^{-} \to \gamma_{\mathrm{ISR}} J/\psi$. In the non-$D\bar{D}$ inclusive MC sample, there are eight exclusive decays, including $\psi(3770) \to J/\psi\pi^{+}\pi^{-},~J/\psi\pi^{0}\pi^{0},~J/\psi\eta,~\phi\eta,~e^{+}e^{-}$, and $\gamma\chi_{c0, c1, c2}$. All particle decays are modelled with {\sc evt}{\sc gen}~\cite{evtgen1,evtgen2} using branching fractions either taken from the PDG~\cite{Ref_pdg}, when available, or otherwise estimated with {\sc lundcharm}~\cite{lundcharm1,lundcharm2}. Final state radiation (FSR) from charged final state particles is incorporated using the {\sc photos} package~\cite{fsr}. The signal decay channels $\psi(3770) \to \gamma \eta_{c}(1S, 2S)$ are generated with an angular distribution proportional to (1 + $ \cos^{2}\theta_{\gamma}$), where $\theta_{\gamma}$ is the polar angle of the radiative photon in the rest frame of the $\psi(3770)$ decay. The $\eta_{c}(1S, 2S) \to K^{0}_{S} K^{\pm} \pi^{\mp}$ decays are generated uniformly in phase space (PHSP). The $\psi(3770) \to \gamma\eta_{c}(1S, 2S)$ decays are denoted as the $\eta_{c}(1S)$ and $\eta_{c}(2S)$ channels, respectively.

\section{EVENT SELECTION}
\label{sec:selection}

Charged tracks detected in the MDC are required to be within the polar angle ($\theta$) range $|\cos\theta|<0.93$. The angle $\theta$ is defined with respect to the $z$-axis, which is the symmetry axis of the MDC. For charged tracks not originating from $K^{0}_{S}$ decays, the distance of the closest approach to the interaction point (IP) along the $z$-axis, $|V_{z}|$, must be less than 10 cm, and that in the transverse plane, $V_{xy}$, must be less than 1 cm. Particle identification (PID) for charged tracks combines measurements of the energy deposited in the MDC (d$E/$d$x$) and the flight time in the TOF to form likelihoods $\mathcal{L}(h)~(h = p, K, \pi)$ for each hadron hypothesis. Tracks are identified as kaons if $\mathcal{L}(K) > \mathcal{L}(\pi)$ and $\mathcal{L}(K) > \mathcal{L}(p)$, and as pions if $\mathcal{L}(\pi) > \mathcal{L}(K)$ and $\mathcal{L}(\pi) > \mathcal{L}(p)$. No PID is applied to charged pions originating from $K^{0}_{S}$ decays.

Photon candidates are identified using isolated showers
in the EMC. The deposited energy of each shower is required to be at least 25 MeV in the barrel region $(|\cos\theta|<0.80)$ and 50 MeV in the end cap region $(0.86<|\cos\theta|<0.92)$. To exclude showers originating from charged tracks, the angle subtended by the EMC shower and the position of the closest charged track at the EMC must be greater than $10^\circ$ as measured from the IP. To suppress electronic noise and energy depositions unrelated to the event, the EMC cluster timing from the reconstructed event start time is required to be within [0, 700] ns. 

$K^{0}_{S}$ candidates are reconstructed from pairs of oppositely charged tracks without $|V_{z}|$ or $V_{xy}$ requirements. A common vertex fit is performed for each oppositely charged track pair under the assumption that both tracks are pions~\cite{Ref_vertexfit}. The decay length from the secondary vertex fit is required to be greater than twice the vertex resolution away from the IP. Among multiple $K^{0}_{S}$ candidates, the one with an invariant mass closest to the known $K^{0}_{S}$ mass~\cite{Ref_pdg} is retained if the mass lies within $\pm 10$ MeV$/c^{2}$ of the nominal value. The fitted $K^{0}_{S}$ candidate parameters are then used as inputs to the subsequent kinematic fit.

In the selection of \(\psi(3770) \to \gamma K^{0}_{S}K^{\pm}\pi^{\mp}\) events, candidates must contain at least four charged tracks and at least one good photon. After reconstructing a $K^{0}_{S}$ candidate, the remaining charged tracks in the event, excluding $\pi^{+}\pi^{-}$ from the $K^{0}_{S}$ decay, must be exactly two with zero net charge, identified as a kaon and a pion by PID. A four-constraint (4C) kinematic fit is performed for the $\gamma \pi^{+}\pi^{-}K^{\pm}\pi^{\mp}$ combination under the $\psi(3770) \to \gamma K^{0}_{S}K^{\pm}\pi^{\mp} \to \gamma \pi^{+}\pi^{-}K^{\pm}\pi^{\mp}$ hypothesis. If multiple photon candidates satisfy the selection, the one with the smallest $\chi^2_{\mathrm{4C}}$ is kept for further analysis. The $\chi_{\mathrm{4C}}^2$ requirement is obtained by optimizing the Punzi figure-of-merit ($\mathrm{FOM} = \frac {\epsilon}{a/2+\sqrt {B}}$)~\cite{Ref_punzi}, where $\epsilon$ denotes the signal efficiency in the $\eta_{c}(1S)$ signal region ([2.7, 3.2] GeV/$c^{2}$) or the $\eta_{c}(2S)$ signal region ([3.45, 3.71] GeV/$c^{2}$) obtained from signal MC samples, $a=3$ is the target significance value, and $B$ denotes the number of background events in the $\eta_{c}(1S, 2S)$ signal regions from inclusive MC samples. Events with $\chi^2_{\mathrm{4C}}<30$ ($\chi^2_{\mathrm{4C}}<35$) are accepted as $\gamma K^{0}_{S}K^{\pm}\pi^{\mp}$ candidates in the $\eta_{c}(1S)$ ($\eta_{c}(2S)$) signal region. To suppress backgrounds from the $\psi(3770) \to \pi^{+}\pi^{-}K^{\pm}\pi^{\mp}$ and $\psi(3770) \to \pi^{0}\pi^{+}\pi^{-}K^{\pm}\pi^{\mp}$ decays, 4C kinematic fits are also performed under zero- and two-photon hypotheses. Candidate events are required to satisfy $\chi^2_{\mathrm{4C}}(\gamma\pi^{+}\pi^{-}K^{\pm}\pi^{\mp}) < \chi^2_{\mathrm{4C}}(\pi^{+}\pi^{-}K^{\pm}\pi^{\mp})$ and $\chi^2_{\mathrm{4C}}(\gamma\pi^{+}\pi^{-}K^{\pm}\pi^{\mp}) < \chi^2_{\mathrm{4C}}(\gamma\gamma\pi^{+}\pi^{-}K^{\pm}\pi^{\mp})$, where the $\chi^2_{\mathrm{4C}}(\gamma\gamma\pi^{+}\pi^{-}K^{\pm}\pi^{\mp})$ are obtained by looping over all photon combinations.

\section{DATA ANALYSIS}

\subsection{Background estimation}

Analysis of the inclusive MC samples using the TopoAna tool~\cite{Ref_topo} indicates that the dominant background contributions come from four sources: (1) $\psi(3770) \to D^{0}\Bar{D}^{0},~(D^{0} \to \pi^{+}K^{-}, \Bar{D}^{0} \to \pi^{0}K_{S}^{0}) + c.c.$; (2) $e^{+}e^{-} \to \pi^{0}K^{0}_{S}K^{\pm}\pi^{\mp}$ with the $\pi^{0}$ decaying into a $\gamma\gamma$ pair; (3) $e^{+}e^{-} \to K^{0}_{S}K^{\pm}\pi^{\mp}$ with a fake photon, ISR photon, or FSR photon; (4) $e^{+}e^{-} \to \gamma_{\mathrm{ISR}}\psi(3686) \to \gamma_{\mathrm{ISR}}\gamma\eta_{c}(1S,2S)$. The few remaining background events are flatly distributed in the $K^{0}_{S}K^{\pm}\pi^{\mp}$ invariant mass spectrum.

Background events from $\psi(3770) \to D^{0}\Bar{D}^{0}$, $(D^{0} \to \pi^{+}K^{-}, \Bar{D}^{0} \to \pi^{0}K_{S}^{0}) + c.c.$ are suppressed by requiring $|\rm{M}(\it K^{\pm}\it\pi^{\mp}) - \it{m}_{\it D^{\rm 0}}| >$ 25 MeV$/c^{2}$, where $m_{\it D^{\rm 0}}$ denotes the known $D^{0}$ mass~\cite{Ref_pdg}.

Background events from $e^{+}e^{-} \to \pi^{0}K^{0}_{S}K^{\pm}\pi^{\mp}$ contaminate the signal through a missing soft photon from the $\pi^{0}$ decay, and the contribution from this process is estimated using a data-driven method~\cite{Ref_datadrive}. The background is measured from data by reconstructing $\gamma\gamma K^{0}_{S}K^{\pm}\pi^{\mp}$ events, where the $\gamma\gamma$ pair forms a $\pi^{0}$ candidate. The selection criteria are the same as those in the signal selection but with an additional photon. The yield of this background in each M$(K^{0}_{S}K^{\pm}\pi^{\mp})$ interval is extracted by fitting the $\pi^{0}$ signal in the M$(\gamma\gamma)$ spectrum. A PHSP MC sample of $e^{+}e^{-} \to \pi^{0}K^{0}_{S}K^{\pm}\pi^{\mp}$ is generated to obtain the relative efficiency between the $\pi^{0}K^{0}_{S}K^{\pm}\pi^{\mp}$ and $\gamma K^{0}_{S}K^{\pm}\pi^{\mp}$ selection criteria in each M$(K^{0}_{S}K^{\pm}\pi^{\mp})$ interval. The background contribution is estimated by 
\begin{equation}\label{pi0bkg}
N_{i}^{\mathrm{bkg}} = N_{i}^{\mathrm{fit}} \times \frac{\epsilon^{\gamma K^{0}_{S}K^{\pm}\pi^{\mp}}}{\epsilon^{\pi^{0}K^{0}_{S}K^{\pm}\pi^{\mp}}},
\end{equation}
where $N_{i}^{\mathrm{bkg}}$ is the number of background events in the $i$-th M$(K^{0}_{S}K^{\pm}\pi^{\mp})$ interval of the $\gamma K^{0}_{S}K^{\pm}\pi^{\mp}$ channel, $N_{i}^{\mathrm{fit}}$ is the fitted yield of background in the $i$-th M$(K^{0}_{S}K^{\pm}\pi^{\mp})$ interval of the $\pi^{0}K^{0}_{S}K^{\pm}\pi^{\mp}$ channel, and $\epsilon^{\gamma K^{0}_{S}K^{\pm}\pi^{\mp}}$ and $\epsilon^{\pi^{0}K^{0}_{S}K^{\pm}\pi^{\mp}}$ are the efficiencies as functions of the $K^{0}_{S}K^{\pm}\pi^{\mp}$ invariant mass with which the $e^{+}e^{-} \to \pi^{0}K^{0}_{S}K^{\pm}\pi^{\mp}$ MC simulated events pass the $\gamma K^{0}_{S}K^{\pm}\pi^{\mp}$ and $\pi^{0}K^{0}_{S}K^{\pm}\pi^{\mp}$ selections, respectively.

The process $e^{+}e^{-} \to K^{0}_{S}K^{\pm}\pi^{\mp}(\gamma_{\mathrm{ISR}}/\gamma_{\mathrm{FSR}})$ can contaminate the signal processes by a fake photon, ISR photon, or FSR photon. To estimate their contributions, an exclusive MC sample is generated for the process $e^{+}e^{-} \to K^{0}_{S}K^{\pm}\pi^{\mp}(\gamma_{\mathrm{ISR}}/\gamma_{\mathrm{FSR}})$ using ConExc~\cite{Ref_conexc}. This event generator is constructed by modeling ISR up to the next leading order correction, and the FSR photon is generated by the {\sc photos} package~\cite{fsr}. The experimental Born cross sections of $e^{+}e^{-} \to K^{0}_{S}K^{\pm}\pi^{\mp}$ from the BaBar Collaboration~\cite{Ref_crosssection} are used as input to the generator. The number of background events is estimated by 
\begin{equation}\label{fsrbkg}
N_{K^{0}_{S}K^{\pm}\pi^{\mp}}^{\mathrm{bkg}} = {\textstyle{\sigma(s) \times \mathcal{L}} \times \mathrm{\epsilon} \times \mathcal{B}},
\end{equation}
where $\sigma(s)$ is the cross section of this process at $\sqrt{s} = 3.773$ GeV~\cite{Ref_crosssection}, $\mathcal{L}$ is the integrated luminosity of the data sample, $\epsilon$ is the detection efficiency from the MC sample, and $\mathcal{B}$ denotes the branching fraction of $K^{0}_{S} \to \pi^{+}\pi^{-}$~\cite{Ref_pdg}. In the subsequent fit, the yield of this background is fixed to the value obtained from Eq.~\ref{fsrbkg} and its shape is modeled using the distribution derived from MC simulation.

Background events from the radiative tail of the $\psi(3686)$ resonance, i.e., $e^{+}e^{-} \to \gamma_{\mathrm{ISR}}\psi(3686) \to \gamma_{\mathrm{ISR}}\gamma\eta_{c}(1S,2S)$, contribute to the peaking backgrounds within the $\eta_{c}(1S,2S)$ signal regions. The number of these backgrounds events is estimated by
\begin{equation}\label{tailbkgeq}
N_{\mathrm{\psi(3686)}}^{\mathrm{bkg}} = {\textstyle{\sigma(s) \times \mathcal{L}} \times \mathrm{\epsilon} \times \prod_{i}\mathcal{B}_{i}},
\end{equation}
where $N_{\mathrm{\psi(3686)}}^{\mathrm{bkg}}$ is the number of background events, $\sigma(s)$ is the cross section of $\psi(3686)$ production at $\sqrt{s} = 3.773~\rm{GeV}$, $\mathcal{L}$ is the integrated luminosity of the data sample, $\epsilon$ is the detection efficiency estimated from the MC sample, and $\mathcal{B}_{i}$ denotes the branching fractions for the intermediate resonance decays: $\psi(3686) \to \gamma \eta_{c}(1S, 2S)$, and $\eta_{c}(1S, 2S) \to K^{0}_{S} K^{\pm} \pi^{\mp}$~\cite{Ref_pdg}. The cross section $\sigma(s)$ can be expressed as 
\begin{equation}\label{crosssection}
\mathrm{\sigma(\it s)} = \int_{0}^{x_{\mathrm{cut}}} W(s,x) \cdot BW(s^{'}(x)) \cdot F_{X}(s^{'}(x))dx,
\end{equation}
where $x$ is the scaled radiated energy in $e^{+}e^{-} \to \gamma_{\mathrm{ISR}}\psi(3686)~(x = 2E_{\gamma_{\mathrm{ISR}}}/\sqrt{s})$, $s^{'}(x) = s(1-x)$ is the squared mass with which the $\psi(3686)$ is produced, $W(s,x)$ is the ISR $\gamma$-emission probability~\cite{Ref_isrpro}, and $BW(s^{'}(x)) = 12\pi\Gamma_{R}\Gamma_{ee}/[(s^{'}-m_{R}^{2})^{2}+m^{2}_{R}\Gamma^{2}_{R}]$ is the relativistic Breit-Wigner formula describing the $\psi(3686)$ resonance. The known $\psi(3686)$ mass $(m_{R})$, width $(\Gamma_{R})$ and $e^{+}e^{-}$ width $(\Gamma_{ee})$ are taken from the PDG~\cite{Ref_pdg}. The term $F_{X}(s^{'}(x)) = (E_{\gamma}(s^{'})/E_{\gamma}(m^{2}_{R}))^{3}$ is the phase-space factor for the $\psi(3686)$ produced with invariant mass $\sqrt{s^{'}}$ relative to $m_{R}$, in which $E_{\gamma}$ is the energy of the transition photon in $\psi(3686) \to \gamma\eta_{c}(1S,2S)$ decays. The threshold cutoff $x_{\mathrm{cut}} = 1 - m^{2}_{X}/s$ is chosen as the upper limit on integration in the definition of $\sigma(s)$, where $m_{X}$ is the nominal mass of the $\eta_{c}(1S,2S)$~\cite{Ref_pdg}. The estimated background events are $N_{\psi(3686)}^{\mathrm{bkg}} = 16.6 \pm 0.2$ for the $\eta_{c}(1S)$ channel and $N_{\psi(3686)}^{\mathrm{bkg}} = 0.19 \pm 0.14$ for the $\eta_{c}(2S)$ channel. 

\subsection{Fit method}

The yields of $\psi(3770) \to \gamma \eta_{c}(1S, 2S) \to \gamma K^{0}_{S}K^{\pm}\pi^{\mp}$ are determined via an unbinned maximum-likelihood fit on the M$(K^{0}_{S}K^{\pm}\pi^{\mp})$ spectrum from data. The signal probability density function (PDF) for the $\eta_{c}(1S,2S)$ state~\cite{Ref_datadrive, Ref_e7} is modeled as
\begin{equation}\label{lineshape}
\textstyle{(E_{\gamma}^{7} \times {BW(m)} \times f_{d}(E_{\gamma}) \times {\epsilon(m)}) \otimes G(\delta m,\sigma)},
\end{equation}
where $E_{\gamma}$ = $\frac{m_{\psi(3770)}^{2}-\rm{M}^{2}}{2m_{\psi(3770)}}$ is the energy of the transition photon in the $\psi(3770)$ meson rest frame, M is the $K^{0}_{S}K^{\pm}\pi^{\mp}$ invariant mass, $m_{\psi(3770)}$ is the $\psi(3770)$ meson nominal mass~\cite{Ref_pdg}. The $BW(m)$ term is a Breit-Wigner function with $\eta_{c}(1S,2S)$ resonance parameters fixed to the PDG values~\cite{Ref_pdg}, while $\epsilon(m)$ is the mass-dependent detection efficiency derived from the MC simulated sample.
Following the convention of charmonium radiative transitions, the $E_\gamma^7$ dependence arises from the dominant $E2$ and $M3$ multipole mixture, as established in Ref.~\cite{Ref_LRC}. For an emitted photon carrying total angular momentum number $l$, the power of $E_\gamma$ is $2l+1$. To suppress the unphysical divergence at low $E_\gamma$, a damping factor $f_{d}(E_{\gamma})$ is adopted from the KEDR experiment~\cite{Ref_damp}:
\begin{equation}\label{kedr}
\textstyle {f_d(E_{\gamma})} = \displaystyle{ {E_{0}^{2}} \over {E_{\gamma}E_{0}+(E_{\gamma}-E_{0})^{2}} },
\end{equation}
where $E_{0}$ = ${m^{2}_{\psi(3770)}-m^{2}_{\eta_{c}(1S,2S)}} \over {2m_{\psi(3770)}}$ denotes the peak energy of the transition photon, and $m_{\eta_{c}(1S,2S)}$ is the nominal mass of the $\eta_{c}(1S,2S)$~\cite{Ref_pdg}. 
The factor $G(\delta m,\sigma)$ accounts for the mass shift and the difference of detector resolution between data and MC simulation. 
The resolution parameters $\delta m$ and $\sigma$ in the Gaussian function are calibrated using dedicated control samples $e^{+}e^{-} \to \gamma_{\mathrm{ISR}} J/\psi \to \gamma_{\mathrm{ISR}} K^{0}_{S}K^{\pm}\pi^{\mp}$ for the $\eta_{c}(1S)$ signal, and $\psi(3770) \to \gamma\chi_{c1} \to \gamma K^{0}_{S}K^{\pm}\pi^{\mp}$ for the $\eta_{c}(2S)$ signal. Both control samples are analyzed with the identical event-selection criteria applied to the signal modes.

In the $\eta_{c}(1S)$ signal region, the fit includes five background components: $e^{+}e^{-} \to \pi^{0}K^{0}_{S}K^{\pm}\pi^{\mp}, e^{+}e^{-} \to K^{0}_{S}K^{\pm}\pi^{\mp}(\gamma_{\mathrm{ISR}}/\gamma_{\mathrm{FSR}})$, $e^{+}e^{-} \to \gamma_{\mathrm{ISR}}\psi(3686) \to \gamma_{\mathrm{ISR}}\gamma\eta_{c}(1S)$, $e^{+}e^{-} \to \gamma_{\mathrm{ISR}} J/\psi$, and other flat backgrounds. The contributions of the first three components are described above. The lineshapes of $e^{+}e^{-} \to K^{0}_{S}K^{\pm}\pi^{\mp}(\gamma_{\mathrm{ISR}}/\gamma_{\mathrm{FSR}})$ and $e^{+}e^{-} \to \gamma_{\mathrm{ISR}}\psi(3686) \to \gamma_{\mathrm{ISR}}\gamma\eta_{c}(1S)$ are described by the MC simulated shapes. The line shape of $e^{+}e^{-} \to \gamma_{\mathrm{ISR}} J/\psi$ is described by the MC simulated shape convolved with a Gaussian function with free parameters, and its yield is floated. Other flat backgrounds are modeled by a second-order polynomial function with floating yield. 

In the $\eta_{c}(2S)$ signal region, the fit includes six background components: $e^{+}e^{-} \to \pi^{0}K^{0}_{S}K^{\pm}\pi^{\mp}$, $e^{+}e^{-} \to K^{0}_{S}K^{\pm}\pi^{\mp}(\gamma_{\mathrm{ISR}}/\gamma_{\mathrm{FSR}})$, $e^{+}e^{-} \to \gamma_{\mathrm{ISR}}\psi(3686) \to \gamma_{\mathrm{ISR}}\gamma\eta_{c}(2S)$, $e^{+}e^{-} \to \gamma_{\mathrm{ISR}} \psi(3686) \to \gamma_{\mathrm{ISR}} K^{0}_{S}K^{\pm}\pi^{\mp}$, $\psi(3770) \to \gamma\chi_{c1}$, and other flat backgrounds. The contributions of the first three components are described above. The line shapes of $e^{+}e^{-} \to K^{0}_{S}K^{\pm}\pi^{\mp}(\gamma_{\mathrm{ISR}}/\gamma_{\mathrm{FSR}})$ and $e^{+}e^{-} \to \gamma_{\mathrm{ISR}}\psi(3686) \to \gamma_{\mathrm{ISR}}\gamma\eta_{c}(2S)$ are described by the MC simulated shapes. The line shape of $e^{+}e^{-} \to \gamma_{\mathrm{ISR}} \psi(3686) \to \gamma_{\mathrm{ISR}} K^{0}_{S}K^{\pm}\pi^{\mp}$ is described by the MC simulated shape convolved with a Gaussian function with free parameters, and its yield is floated. The line shape of $\psi(3770) \to \gamma\chi_{c1}$ is modeled with the same form as Eq.~\ref{lineshape}, and its yield is floated. Other flat backgrounds are described by a second-order polynomial function with floating yield.

\subsection{Numerical results}

\begin{figure*}[htb]
\begin{center}
\includegraphics[width=.48\textwidth]{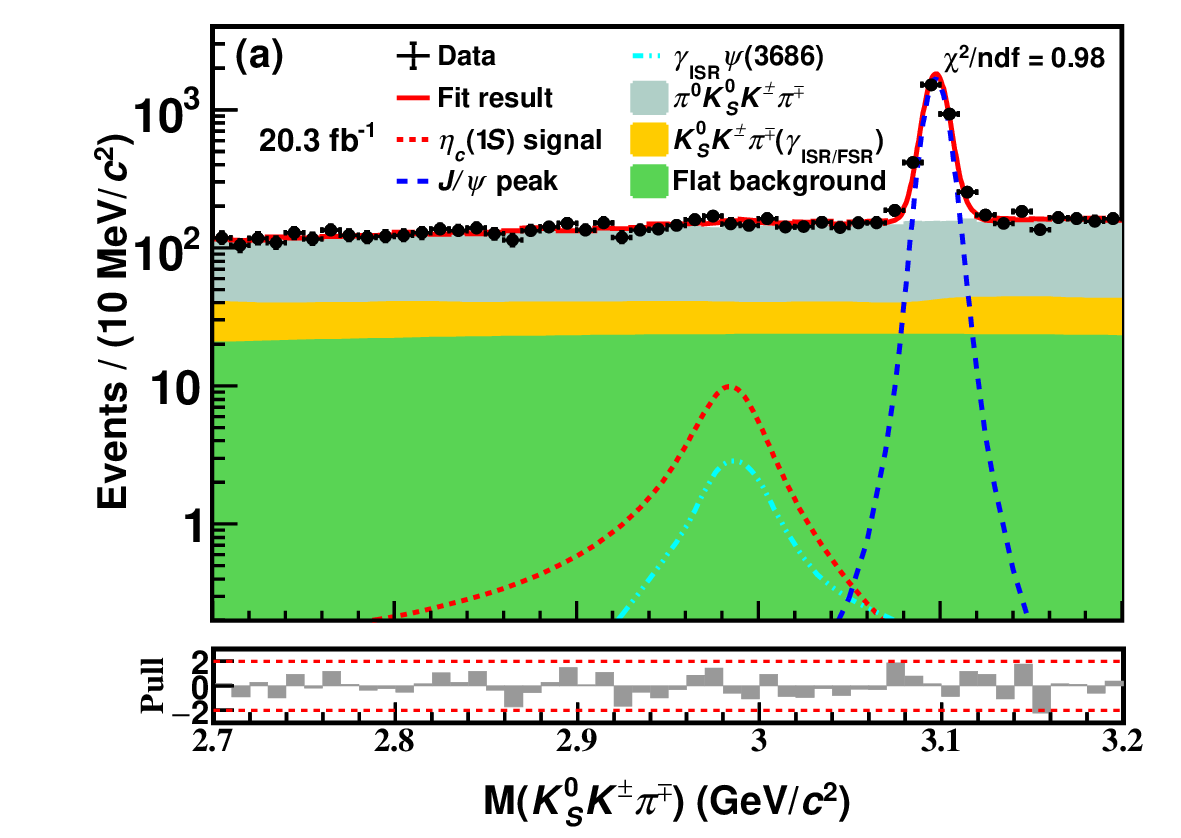}
\includegraphics[width=.48\textwidth]{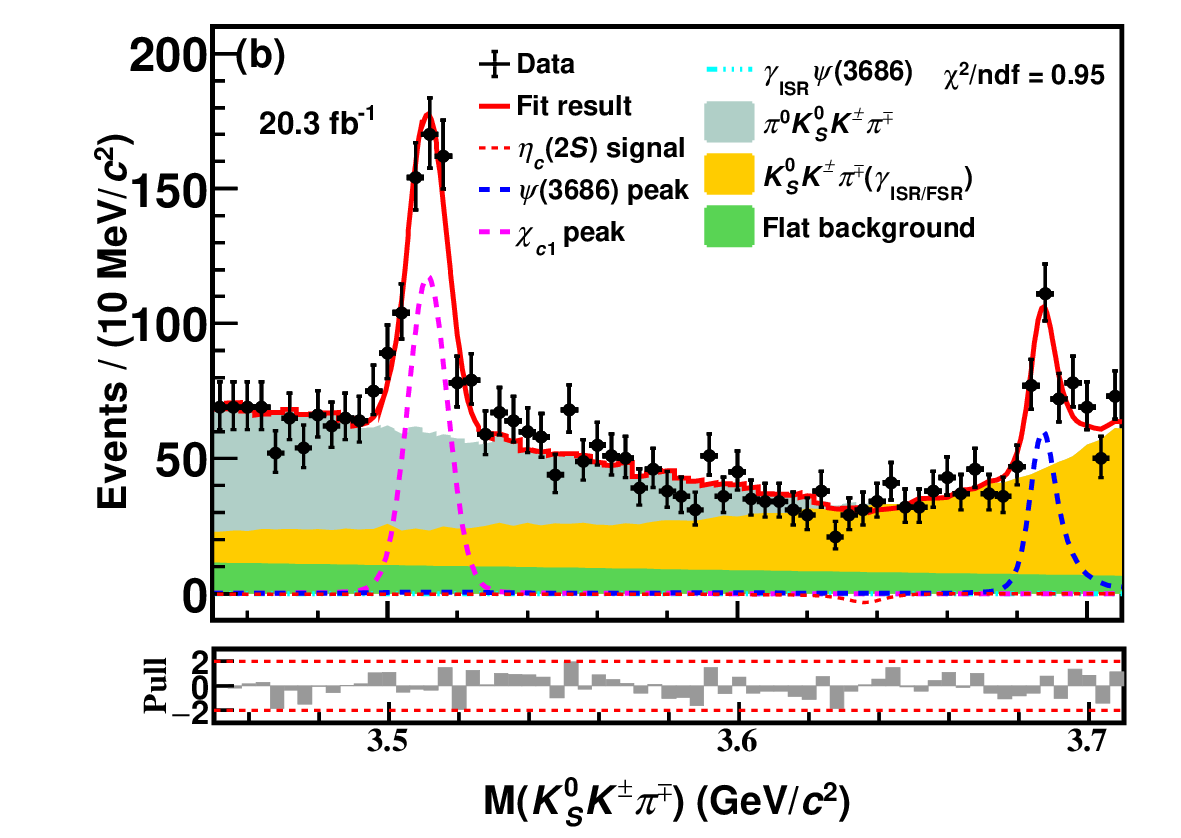}
\caption{The $K^{0}_{S}K^{\pm}\pi^{\mp}$ invariant mass spectra from data and the related fit results in the (a) $\eta_{c}(1S)$ and (b) $\eta_{c}(2S)$ signal regions.}
\label{massfit}
\end{center}
\end{figure*}

Unbinned maximum likelihood fits to the $K^{0}_{S}K^{\pm}\pi^{\mp}$ invariant mass spectra are performed separately in the $\eta_{c}(1S)$ and $\eta_{c}(2S)$ signal regions. The fit results are shown in Figs.~\ref{massfit}(a) and~\ref{massfit}(b), with goodness of fit values of $\chi^{2}/\mathrm{ndf} = 42.2/43$ and $55.2/58$, respectively, where ndf denotes the degrees of freedom. The extracted signal yields are $N_{\eta_{c}(1S)} = 53 \pm 28$ and $N_{\eta_{c}(2S)} = -25 \pm 13$. The signal significances are estimated using $-2\rm{ln}(\mathcal{L}_{0}/\mathcal{L}_{\rm{max}})$, where $\mathcal{L}_{0}$ and $\mathcal{L}_{\rm{max}}$ are the maximized likelihoods without and with $\eta_{c}(1S,2S)$ signals, respectively, taking into account the differences in the number of degrees of freedom ($\Delta \rm{ndf}$)~\cite{Ref_signi}. The resulting statistical significances of the $\eta_{c}(1S,2S)$ signals are 1.1$\sigma$ and 1.0$\sigma$, respectively. Since no significant signal is observed for either the $\eta_{c}(1S)$ or $\eta_{c}(2S)$, 90$\%$ C.L. upper limits on the number of signal events are computed, $N_{\mathrm{C.L.}(90\%)}$, by solving the equation~\cite{Ref_Bayesian},
\begin{equation}\label{uplimit}
\int^{N_{\mathrm{C.L.}(90\%)}}_{0}{\mathcal{L}(n)}dn/\int^{+\infty}_{0}{\mathcal{L}(n)}dn = 0.90,
\end{equation}
where $n$ is the assumed number of signal events, and $\mathcal{L}(n)$ is the corresponding maximized profiled likelihood of the fit with each fixed value of $n$. The resulting 90$\%$ C.L. upper limits on the number of signal events, including systematic uncertainties (discussed below), are $N_{\mathrm{C.L.}(90\%)}<139$ for the $\eta_{c}(1S)$ channel and $N_{\mathrm{C.L.}(90\%)}<156$ for the $\eta_{c}(2S)$ channel.

	
\section{Systematic Uncertainties}\label{sec:sysU}

The systematic uncertainties are classified into two categories: additive and multiplicative terms. The systematic uncertainties from the masses and widths of the $\eta_{c}(1S,2S)$ and the fit model are additive. The other sources of systematic uncertainties are multiplicative, and are listed in Table~\ref{sys}. The multiplicative terms include tracking efficiency, $K^{0}_{S}$ reconstruction, photon reconstruction, kinematic fit, integrated luminosity of the $\psi(3770)$ data sample, cross section of $e^+e^-\to\psi(3770)$, $D^{0}$ veto, branching fractions, and MC statistics.

The tracking efficiency differences for $K^{\pm}$ and $\pi^{\pm}$ between data and MC simulations are studied with the control samples $\psi(3770) \to D^{0}\bar{D}^{0}(D^{+}D^{-})$, with $D^{0} \to K^{-}\pi^{+}, K^{-}\pi^{+}\pi^{+}\pi^{-}$ and $\bar{D}^{0} \to K^{+}\pi^{-}, K^{+}\pi^{-}\pi^{-}\pi^{+}$ ($D^{+} \to K^{-}\pi^{+}\pi^{+}$ and $D^{-} \to K^{+}\pi^{-}\pi^{-}$). The uncertainties are estimated to be 0.9$\%$ for the $\eta_{c}(1S)$ channel, and 1.0$\%$ for the $\eta_{c}(2S)$ channel.

The $K^{0}_{S}$ reconstruction uncertainty is determined using the control samples $\psi(3770) \to D^{0}\bar{D}^{0}(D^{+}D^{-})$, with $D^{0} \to K^{-}\pi^{+}, K^{-}\pi^{+}\pi^{+}\pi^{-}, K^{-}\pi^{+}\pi^{0}$ and $\bar{D}^{0} \to K_{S}^{0}\pi^{+}\pi^{-}, K_{S}^{0}\pi^{0}, K_{S}^{0}\pi^{+}\pi^{-}\pi^{0}$ ($D^{+} \to K^{-}\pi^{+}\pi^{+}$ and $D^{-} \to K_{S}^{0}\pi^{-}, K_{S}^{0}\pi^{-}\pi^{0}, K_{S}^{0}\pi^{+}\pi^{-}\pi^{-}$), and the uncertainties are estimated to be 0.4$\%$ for the $\eta_{c}(1S)$ channel, and 0.3$\%$ for the $\eta_{c}(2S)$ channel.

The uncertainty on the photon reconstruction efficiency is estimated to be $1\%$ per photon, based on studies of the control samples $J/\psi \to \rho^{0}\pi^{0}$ and $e^{+}e^{-} \to \gamma\gamma$~\cite{Ref_photon}.

The differences in the $\chi^2_{\mathrm{4C}}$ distributions between data and MC simulations are estimated by correcting the track helix parameters~\cite{Ref_helixcor}. The correction factors are extracted from the control samples $\psi(3686) \to \gamma\chi_{c0} \to \gamma 3(\pi^{+}\pi^{-})$ and $\psi(3686) \to \gamma\chi_{c0} \to\gamma 2(K^{+}K^{-})$. The uncertainties are estimated to be 0.5$\%$ for the $\eta_{c}(1S)$ channel, and 1.3$\%$ for the $\eta_{c}(2S)$ channel.

The integrated luminosity of the $\psi(3770)$ resonance data sample is determined to be $(20.3 \pm 0.08)~\rm{fb}^{-1}$~\cite{Ref_lum1,Ref_lum2}, corresponding to an uncertainty of 0.4$\%$.

To determine the total number of $\psi(3770)$ mesons, we use the $e^+e^-\to\psi(3770)$ Born-level cross section ($\sigma^{0}_{\psi(3770)}$) at $\sqrt{s} = 3.773~\rm{GeV}$, which is calculated using a relativistic Breit-Wigner formula with fixed $\psi(3770)$ resonance parameters~\cite{Ref_pdg}. The uncertainty of the cross section is $7.3\%$, which is dominantly due to uncertainty in the $\psi(3770)$ resonance parameters.

The uncertainty related to the $D^{0}$ veto is estimated using the control sample $e^{+}e^{-} \to \gamma_{\mathrm{ISR}} J/\psi \to \gamma_{\mathrm{ISR}} K^{0}_{S} K^{\pm} \pi^{\mp}$. The efficiency difference between data and MC simulation is determined to be 1$\%$.

The branching fractions of $K^{0}_{S} \to \pi^{+}\pi^{-}$, $\eta_{c}(1S) \to K^{0}_{S} K^{\pm} \pi^{\mp}$, and $\eta_{c}(2S) \to K^{0}_{S} K^{\pm} \pi^{\mp}$ are $(69.20 \pm 0.05)\%$~\cite{Ref_pdg}, $(2.60 \pm 0.21 \pm 0.20)\%$~\cite{Ref_bf1}, and $(0.60^{+0.38}_{-0.32})\%$~\cite{Ref_bf2,Ref_bf3}, respectively. The corresponding uncertainties are 11$\%$ for the $\eta_{c}(1S)$ channel and 63$\%$ for the $\eta_{c}(2S)$ channel.

The uncertainty due to MC statistics is estimated to be $0.4\%$ by assuming a binomial distribution for the detection efficiency of the signal MC.

To estimate the uncertainty associated with the signal simulation model, the intermediate states $K_{0}^{*}(1430)$ and $K_{2}^{*}(1430)$ are considered following the Dalitz-plot analysis of the $\eta_{c}(1S,2S) \to K^{0}_{S} K^{\pm} \pi^{\mp}$ decays by the LHCb collaboration~\cite{LHCb:2023evz}. Mixed MC samples are generated for these two processes, and the differences in efficiency with respect to the nominal signal MC samples are evaluated. The observed difference is taken as the related uncertainty.

The additive systematic uncertainties are evaluated by varying the fit model and $\eta_{c}(1S,2S)$ parameters: (1) the uncertainty related to the $\eta_{c}(1S,2S)$ resonance parameters is estimated by varying the parameters by one standard deviation from the PDG values~\cite{Ref_pdg}; (2) the uncertainty related to the photon energy dependence is estimated by replacing $E_{\gamma}^{7}$ with $E_{\gamma}^{5}$~\cite{Ref_LRC}; (3) the uncertainty related to the damping factor is estimated by removing it from the fit; (4) the uncertainty related to the Gaussian resolution function is estimated by varying the parameters by one standard deviation; (5) the uncertainties related to the $\pi^{0}K^{0}_{S}K^{\pm}\pi^{\mp}$, $K^{0}_{S}K^{\pm}\pi^{\mp}(\gamma_{\mathrm{ISR}}/\gamma_{\mathrm{FSR}})$, and $\gamma_{\mathrm{ISR}}\psi(3686)$ background yields are estimated by changing their magnitudes by one standard deviation; (6) the uncertainty related to the shape of the $e^{+}e^{-} \to K_{S}^{0} K^{\pm}\pi^{\mp}(\gamma_{\mathrm{ISR}}/\gamma_{\mathrm{FSR}})$ background is estimated by varying the smooth parameters of the MC simulated shape; (7) the uncertainties related to the shapes of the $e^{+}e^{-} \to \gamma_{\mathrm{ISR}}\psi(3686) \to \gamma_{\mathrm{ISR}}\gamma\eta_{c}(1S, 2S)$, and $e^{+}e^{-} \to \gamma_{\mathrm{ISR}} J/\psi/\psi(3686) \to \gamma_{\mathrm{ISR}}K^{0}_{S}K^{\pm}\pi^{\mp}$ backgrounds are estimated by replacing the MC simulated shape with a Breit-Wigner function convolved with a single Gaussian function with floating parameters; (8) the uncertainty related to the flat background is estimated by replacing the second-order polynomial with a third-order polynomial.

Systematic uncertainties are incorporated into our calculations of upper limits in two steps. First, in determining the additive systematic uncertainty from the masses and widths of the $\eta_{c}(1S,2S)$ and the fit model (as described above), we calculate the upper limit for each possible fit configuration and determine the most conservative upper limit at 90$\%$ C.L. on the number of signal events. Then, the likelihood with that most conservative upper limit is convolved with a Gaussian function whose width is equal to the corresponding total multiplicative uncertainty~\cite{Ref_uplimit}:
\begin{equation}\label{calupper}
\textstyle {L^{\prime}(N)} = \displaystyle{ \int^{1}_{0} L(\frac{S}{\hat{S}}N)\rm{exp}[-\frac{(\it S-\hat{S})^{\rm 2}}{2\sigma^{2}_{\it S}}]d\it S },
\end{equation}
where ${L^{'}(N)}$ is the convolved likelihood, $L$ is the original likelihood as a function of the number of signal events $N$, $S$ is the expected efficiency under a given systematic variation of the multiplicative sources, $\hat{S}$ is the nominal efficiency obtained from the signal MC sample, and $\sigma_{S}$ is the total multiplicative uncertainty. Here $S$ is treated as a floating parameter representing possible deviations of the efficiency due to systematic sources, and its probability distribution is modeled by the Gaussian term in the integrand. The resulting 90$\%$ C.L. upper limits on the number of signal events are $N_{\mathrm{C.L.}(90\%)}<139$ for the $\eta_{c}(1S)$ channel and $N_{\mathrm{C.L.}(90\%)}<156$ for the $\eta_{c}(2S)$ channel.

\begin {table}[htb]
\begin{center}
{\caption {Relative multiplicative systematic uncertainties ($\%$) on the measurements of the upper limits on $\mathcal{B}(\psi(3770) \to \gamma \eta_{c}(1S,2S))$ at 90$\%$ C.L. The multiplicative uncertainties are assumed to be independent and are added in quadrature to obtain the total multiplicative uncertainty.}
\label{sys}}
\begin {tabular}{l c c c}\hline\hline
Source & $\eta_{c}(1S)$ & $\eta_{c}(2S)$   \\   \hline
Tracking & 0.9 & 1.0  \\
$K^{0}_{S}$ reconstruction  &  0.4 &  0.3 \\
Photon reconstruction  &  1.0 &  1.0 \\	
Kinematic fit &  0.5 &  1.3 \\
Integrated luminosity &  0.4 &  0.4 \\	
Cross section of $e^+e^- \to \psi(3770)$ & 7.3 & 7.3 \\
$D^{0}$ veto   & 1.0 & 1.0 \\
Branching fractions & 11 & 63 \\
MC statistics & 0.4 & 0.4 \\
Intermediate resonance & 0.2 & 1.4\\ \hline
Total multiplicative uncertainty  &   13.3 &  63.5 \\
\hline
\hline
\end{tabular}
\end{center}
\end{table}

{
\renewcommand{\arraystretch}{1.5}
\setlength{\tabcolsep}{4pt}
\begin{table*}[htb]
\begin{center}
\caption{Comparisons of the experimental measurements with theoretical predictions.}
\label{gammasum2}
\begin{tabular}{cccccc}
\hline
\hline
         & This work & Previous measurement~\cite{Ref_3fb} & IML~\cite{Ref_iml} & LQCD~\cite{Ref_lqcd} & Ref.~\cite{Ref_LRC} \\ \hline
        $\Gamma(\psi(3770) \to \gamma \eta_{c}(1S))$~(keV) & $< 5.5$  & $<18.6$ & $17.14^{+22.9}_{-12.03}$ & $10 \pm 11$ & 1.54 \\ \hline
        $\Gamma(\psi(3770) \to \gamma \eta_{c}(2S))$~(keV) & $< 29.4$ & $<54.8$  & $1.82^{+1.95}_{-1.19}$  & - & 0.0233 \\     
         \hline
        \hline
\end{tabular}
\end{center}
\end{table*}
}

\section{upper limits}

The upper limits on the product branching fractions are calculated as
\begin{equation}\label{upbr}
\mathcal{B}_{1}\times\mathcal{B}_{2} < \frac{N_{\mathrm{C.L.}}}{\epsilon\cdot\mathcal{L}\cdot\sigma^{0}_{\psi(3770)}\cdot\frac{1}{|1-\Pi|^{2}}\cdot(1+\sigma)\cdot\mathcal{B}_{3}},
\end{equation}
where $\mathcal{B}_{1}$ and $\mathcal{B}_{2}$ represent $\mathcal{B}(\psi(3770) \to \gamma \eta_{c}(1S,2S))$ and $\mathcal{B}(\eta_{c}(1S,2S) \to K^{0}_{S}K^{\pm}\pi^{\mp})$, respectively, $N_{\mathrm{C.L.}}$ is the upper limit on the number of signal events, $\epsilon$ is the detection efficiency (25.9$\%$ for $\eta_{c}(1S)$ and 23.2$\%$ for $\eta_{c}(2S)$), $\mathcal{L}$ is the integrated luminosity, $\sigma^{0}_{\psi(3770)}$ is the cross section of $\psi(3770)$ production at 3.773 GeV, $\frac{1}{|1-\Pi|^{2}} = 1.056$ is the vacuum polarization factor~\cite{Ref_vacuum}, $(1+\sigma) = 0.714$ is the radiative correction factor~\cite{kkmc,Ref_isrpro,Ref_radcor1}, and $\mathcal{B}_{3}$ is the branching fraction for $K^{0}_{S} \to \pi^{+}\pi^{-}$. The vacuum polarization and radiative correction factors are obtained from the {\sc kkmc} generator. The uncertainties in $\mathcal{B}(\eta_c(1S,2S) \to K^{0}_{S}K^{\pm}\pi^{\mp})$ are not included in the product branching fraction upper limits. The results are $\mathcal{B}(\psi(3770) \to \gamma \eta_{c}(1S)) \times \mathcal{B}(\eta_{c}(1S) \to K^{0}_{S} K^{\pm} \pi^{\mp}) < 5.0 \times 10^{-6}$ at $90\%~\mathrm{C.L.}$ and $\mathcal{B}(\psi(3770) \to \gamma \eta_{c}(2S)) \times \mathcal{B}(\eta_{c}(2S) \to K^{0}_{S} K^{\pm} \pi^{\mp}) < 3.7 \times 10^{-6}$ at $90\%~\mathrm{C.L.}$. Using $\mathcal{B}(\eta_{c}(1S) \to K^{0}_{S} K^{\pm} \pi^{\mp}) = (2.60 \pm 0.21 \pm 0.20)\%$~\cite{Ref_bf1}, $\mathcal{B}(\eta_{c}(2S) \to K^{0}_{S} K^{\pm} \pi^{\mp}) = (0.60^{+0.38}_{-0.32})\%$~\cite{Ref_bf2,Ref_bf3}, and $\Gamma^{\mathrm{tot}}_{\psi(3770)} = 27.2 \pm 1.0$ MeV~\cite{Ref_pdg}, the upper limits on the partial decay widths at $90\%~\rm{\mathrm{C.L.}}$ are
\begin{equation}\label{uplimit2}
\textstyle{
\left\{
\begin{aligned}
\Gamma(\psi(3770) \to \gamma \eta_{c}(1S)) <  5.5~\rm{keV} \\
\Gamma(\psi(3770) \to \gamma \eta_{c}(2S)) < 29.4~\rm{keV}
\end{aligned}
\right.}.
\end{equation}
In the calculation of the partial decay widths, an additional multiplicative uncertainty from $\Gamma^{\mathrm{tot}}_{\psi(3770)}$ (3.7$\%$) is included. 

\section{Summary}

In summary, using a data sample with an integrated luminosity of $20.3~\mathrm{fb}^{-1}$ at $e^+e^-$ center-of-mass energy 3.773 GeV collected with the BESIII detector operating at the BEPCII collider, searches for $\psi(3770) \to \gamma \eta_{c}(1S, 2S)$ via $\eta_{c}(1S, 2S) \to K^{0}_{S} K^{\pm} \pi^{\mp}$ are improved over previous measurements~\cite{Ref_3fb}, but no significant $\eta_{c}(1S, 2S)$ signals are observed. Upper limits on the branching fractions at 90$\%$ C.L. are measured to be $\mathcal{B}(\psi(3770) \to \gamma \eta_{c}(1S)) \times \mathcal{B}(\eta_{c}(1S) \to K^{0}_{S} K^{\pm} \pi^{\mp}) < 5.0 \times 10^{-6}$; $\mathcal{B}(\psi(3770) \to \gamma \eta_{c}(2S)) \times \mathcal{B}(\eta_{c}(2S) \to K^{0}_{S} K^{\pm} \pi^{\mp}) < 3.7 \times 10^{-6}$. The upper limits on the partial decay widths at 90$\%$ C.L. are determined to be $\Gamma(\psi(3770) \to \gamma \eta_{c}(1S)) < 5.5~\rm{keV}$ and $\Gamma(\psi(3770) \to \gamma \eta_{c}(2S)) < 29.4$ keV based on the values of $\mathcal{B}(\eta_c(1S,2S) \to K^{0}_{S}K^{\pm}\pi^{\mp})$~\cite{Ref_bf1,Ref_bf2,Ref_bf3}. Our result is compatible with the previous BESIII measurements~\cite{Ref_3fb}. Using a data sample seven times larger than that used for the previous measurements, the upper limits at 90$\%$ confidence level on $\Gamma(\psi(3770) \to \gamma \eta_{c}(1S))$ and $\Gamma(\psi(3770) \to \gamma \eta_{c}(2S))$ are reduced to approximately 30$\%$ and 54$\%$ of the previous results, respectively.

Table~\ref{gammasum2} compares our results with the previous BESIII measurements~\cite{Ref_3fb} and various theoretical predictions~\cite{Ref_iml,Ref_LRC,Ref_lqcd}. The upper limit on $\Gamma(\psi(3770) \to \gamma \eta_{c}(1S))$ is within the 1$\sigma$ range of the IML~\cite{Ref_iml} and lattice QCD~\cite{Ref_lqcd} calculations, and is larger than the prediction obtained using the relativistic BS equation method that incorporates large relativistic corrections~\cite{Ref_LRC}. For $\Gamma(\psi(3770) \to \gamma \eta_{c}(2S))$, due to the limited statistics of the data sample and the large uncertainty in $\mathcal{B}(\eta_{c}(2S) \to K^{0}_{S} K^{\pm} \pi^{\mp})$, the upper limit is approximately one order of magnitude larger than the IML prediction and significantly larger than the prediction obtained using the relativistic BS equation method that incorporates large relativistic corrections.

	
\section{ACKNOWLEDGMENTS}

The BESIII Collaboration thanks the staff of BEPCII (https://cstr.cn/31109.02.BEPC) and the IHEP computing center for their strong support. This work is supported in part by National Key R\&D Program of China under Contracts Nos. 2025YFA1613900, 2023YFA1606000, 2023YFA1606704, 2025YFA1613900; National Natural Science Foundation of China (NSFC) under Contracts Nos. 11635010, 11935015, 11935016, 11935018, 12025502, 12035009, 12035013, 12061131003, 12192260, 12192261, 12192262, 12192263, 12192264, 12192265, 12221005, 12225509, 12235017, 12342502, 12361141819, 12535005; the Chinese Academy of Sciences (CAS) Large-Scale Scientific Facility Program; the Strategic Priority Research Program of Chinese Academy of Sciences under Contract No. XDA0480600; CAS under Contract No. YSBR-101; 100 Talents Program of CAS; The Institute of Nuclear and Particle Physics (INPAC) and Shanghai Key Laboratory for Particle Physics and Cosmology; Agencia Nacional de Investigaci\'on y Desarrollo de Chile (ANID), Chile under Contract No. ANID CCTVal CIA250027; Istituto Nazionale di Fisica Nucleare, Italy; Knut and Alice Wallenberg Foundation under Contracts Nos. 2021.0174, 2021.0299, 2023.0315; Ministry of Development of Turkey under Contract No. DPT2006K-120470; National Research Foundation of Korea under Contract No. RS-2026-25486791; National Science and Technology fund of Mongolia; Polish National Science Centre under Contract No. 2024/53/B/ST2/00975; STFC (United Kingdom); Swedish Research Council under Contract No. 2019.04595; U. S. Department of Energy under Contract No. DE-FG02-05ER41374


\end{document}